\documentstyle{article}
\title{The Atiyah-Singer index theorem and the gauge field copy problem\footnote{This paper was published in {\em J. Phys. A: Math. Gen.\/} {\bf 30} 5511-5516 (1997).}}
\author{A. S. Sant'Anna\\Department of Mathematics\\Federal University at Paran\'a\\P.O.Box 19.081, 81.531-990, Curitiba, PR, Brazil\and N. C. A. da Costa\\Research Group on Logic and Foundations.\\Institute for Advanced Studies, University of S\~{a}o Paulo.\\Av. Prof. Luciano Gualberto, trav. J, 374.\\05655-010 S\~{a}o Paulo SP Brazil.\and F. A. Doria\\Research Center on\\ Mathematical Theories of Communication\\School of Communications,\\Federal University at Rio de Janeiro.\\Av. Pasteur, 250. 22295-900 Rio RJ Brazil.}
\date{ }

\begin{document}
\maketitle
\newcounter{cms}
\setlength{\unitlength}{1mm}
\begin{abstract}
$K$-theory allows us to define an analytical condition for the existence of `false' gauge field copies through the use of the Atiyah-Singer index theorem. After establishing that result we discuss a possible extension of the same result without the help of the index theorem and suggest possible related lines of work.
\end{abstract}
\section{Introduction}
\newtheorem{definicao}{Definition}[section]
\newtheorem{teorema}{Theorem}[section]
\newtheorem{lema}{Lemma}[section]
\newtheorem{corolario}{Corolary}[section]
\newtheorem{proposicao}{Proposition}[section]
\newtheorem{axioma}{Axiom}[section]
\newtheorem{observacao}{Remark}[section]

	The gauge field copy question is one of those unexpected phenomena that creep up in mathematical physics when we go from linear objects to their nonlinear extensions. Linear gauge fields, say, the electromagnetic field, admit a single potential over a nice neighborhood modulo gauge tranformations. However when we go from the linear to the nonlinear domain, that nice relation between fields and potentials breaks down. (We emphasize the adjective: the relation between potentials and fields in linear gauge fields is a `nice' one because it reflects the very deep $\partial^{2} = 0$ relation in homological algebra and in algebraic topology; for a simple application of that relation to mathematical physics see \cite{Doria-78}.)

	In the nonlinear, nonabelian case, some gauge fields admit two or more potentials which cannot be made equivalent (even locally) modulo gauge transformations. Such an ambiguity is known as the gauge field copy problem and was discovered in 1975 by T.T. Wu and C.N. Yang \cite{Wu}. Gauge copies fall into two cases:
\begin{itemize}
\item{{\em True copies}: Here the gauge field can be derived from at least two different potentials that aren't even locally related by a gauge transformation;}
\item{{\em False copies}: In this case the field can be derived from potentials that are always locally related by a gauge transformation.}
\end{itemize}
For a review of the geometric phenomena behind the copy problem see \cite{Costa-Amaral}.

	Let $P(M,G)$ be a principal fiber bundle, where $M$ is a finite-dimensional smooth real manifold and $G$ is a
finite-dimensional semi-simple Lie group. We denote by $(P,\alpha)$ the principal fiber bundle $P(M,G)$ endowed with the connection-form $\alpha$, and by $L$ the field that corresponds to the potential $A$ associated to $\alpha$. We mean by that that $A$ is a family $\{A_{\lambda}\}_{\lambda\in \Lambda}$ of $l(G)$-valued one-forms, where $l(G)$ is the Lie algebra associated to $G$, and for each $\lambda\in\Lambda$, $A_{\lambda}$ is defined on an open subset $U_{\lambda}$ of $M$ with $D_{A_{\lambda}} = dA_{\lambda}+\frac{1}{2}[A_{\lambda},A_{\lambda}] = L$ on $U_{\lambda}$ and with $M = \bigcup_{\lambda\in\Lambda}U_{\lambda}$. We consider also that $A_{\lambda}$ and $A_{\lambda'}$ are gauge equivalent for $\lambda\neq\lambda'$. We recall that an automorphism of a principal fiber bundle $P(M,G)$ with projection $\pi$ is a diffeomorphism $f:P\rightarrow P$ such that $f(pg) = f(p)g$ for all $g\in G$, $p\in P$. A gauge transformation is an automorphism $f:P\rightarrow P$ such that $\overline{f} = 1_{M}$, where $\overline{f}:M\rightarrow M$ is the diffeomorphism induced by $f$ given by $\overline{f}(\pi(p)) = \pi(f(p))$. Given a principal fibre bundle $P(M,G)$ and a Lie subgroup $G'$ of $G$, we say that $P(M,G)$ is reducible to the bundle $P'(M,G')$ if and only if there is a monomorphism $f':G'\rightarrow G$ and an imbedding $f'':P'\rightarrow P$ such that $f''(u'a') = f''(u')f'(a')$, for all $u'\in P'$ and $a'\in G'$.

\begin{definicao}
\begin{enumerate}
\item{The field $L$ or the potential $A$ are said to be {\bf reducible} if the corresponding bundle $(P,\alpha)$ is reducible.}
\item{If $U\subset M$ is a nonvoid open set then $L$ or $A$ are said to be {\bf locally reducible over} $U$ whenever $(P,\alpha)\mid _{U}$ is reducible.}
\item{$L$ or $A$ are said to be {\bf fully irreducible} if they are not locally reducible.}$\Box$
\end{enumerate}\label{def-redutibilidade}
\end{definicao}

	Our main results are essentially based on a theorem that gives a topological condition for the existence of false copies and on the Atiyah-Singer index theorem. But before we state the topological condition for the existence of false gauge field copies, we find interesting to recall the Ambrose-Singer theorem, since it will be used to derive such a result:

\begin{proposicao}
(Ambrose-Singer) Let $P(M,G)$ be a principal fiber bundle, where $M$ is connected and paracompact. Let $\Gamma$ be a connection in $P$, $L$ the curvature form, $\Phi(u)$ the holonomy group with reference point $u\in P$ and $P(u)$ the holonomy bundle through $u$ of $\Gamma$. Then the Lie algebra of $\Phi(u)$ is equal to the subspace of $l(G)$ spanned by all elements of the form $L_{v}(X,Y)$, where $v\in P(u)$ and $X$ and $Y$ are arbitrary horizontal vectors at $v$.\label{Ambrose-Singer}
\end{proposicao}

	The proof of this theorem (also known as the holonomy theorem), can be found in \cite{Kobayashi}. It is considered that we may assume $P(u) = P$, which means that $\Phi(u) = G$.

	Now we establish the topological condition for the existence of false gauge field copies, based on a result due to Doria \cite{Doria-81}:

\begin{proposicao}
Let $P(M,G)$ be as above together with the fact that $G$ be semi-simple and $M$ is connected and paracompact. $L$ is falsely copied, that is, $L$ has different potentials that are locally related by a gauge transformation if and only if $L$ is reducible.\label{Teo-Doria}
\end{proposicao}

	{\em Proof}: If $L$ is reducible, then $P(M,G)$ is reducible (definition \ref{def-redutibilidade}). This means that $P(M,G)$ can be reduced to a nontrivial $P'(M,G')$, where $G'$ is the Ambrose-Singer holonomy group (it corresponds to the group $\Phi(u)$ in proposition \ref{Ambrose-Singer}). If we assume
\[A_{\mu} = B_{\mu} + \partial_{\mu}h',\] 
where $\partial_{\mu}=_{def}\partial/\partial x^{\mu}$, $x^{\mu}$ is a coordinate of a coordinate system at $U\subset M$ (such that the bundle is trivial over $U$), $B_{\mu}$ takes values in $l(G)$ and $h'$ takes values on $l(G')$, then:
\[F_{\mu\nu}(A) = \partial_{\mu}(B_{\nu}+\partial_{\nu}h') - \partial_{\nu}(B_{\mu} + \partial_{\mu}h') + (B_{\mu} + \partial_{\mu}h')(B_{\nu} + \partial_{\nu}h') - (B_{\nu} + \partial_{\nu}h')(B_{\mu} + \partial_{\mu}h'),\]
where $F_{\mu\nu}(A)$ denotes the components of the curvature form $F$ associated to the connection form $A$.

	Hence:

\[F_{\mu\nu}(A) = F_{\mu\nu}(B) + 0_{field}.\]

	The sufficient condition is proved as follows: if the holonomy group is semi-simple as indicated, and if $A_{\mu} = B_{\mu} + \partial_{\mu}h'$ as indicated, then there is a reducibility $G\oplus G'\rightarrow G'$.$\Box$

	We now suppose that $X$ is a compact smooth manifold and that $G$ is a compact Lie group acting smoothly on $M$. The Atiyah-Singer index theorem can be stated as \cite{Shanahan}:

\begin{proposicao}
Let $\chi$ and $\vartheta$ be complex vector bundles defined over $X$. If $D:C^{\infty}(X;\chi)\rightarrow C^{\infty}(X;\vartheta)$ is a $G$-invariant elliptic partial differential operator on $X$, which sends cross-sections of $\chi$ to cross-sections of $\vartheta$, then {\rm index}$_{G}D = t_{{\rm ind}_{G}^{X}}(\sigma(D))$, where $t_{{\rm ind}}$ is the topological index defined on $K_{G}(TX)$ and $\sigma(D)$ is the symbol of $D$.$\Box$\label{Atiyah-Singer-D}
\end{proposicao}

	(For the proof and notational features see \cite{Shanahan}.) Another version of the index theorem \cite{Atiyah-Index-I} asserts:
\begin{proposicao}
The analytical index $a_{{\rm ind}_{G}}$ and the topological index $t_{{\rm ind}_{G}^{X}}$ coincide as homomorphisms $K_{G}(TX)\rightarrow R(G)$.$\Box$
\end{proposicao}
(Proof in the reference.)

\section{A necessary condition for false copies}

	Gauge fields and gauge potentials can be seen and defined as cross-sections of vector bundles associated to the principal fiber bundle $P(M,G)$. More specifically, potential space (or connection space) coincides with the space of all $C^{k}$ cross-sections of the vector bundle $E$ of $l(G)$-valued 1-forms on $M$, where $l(G)$ is the group's Lie algebra, while field space (or curvature space) coincides with the space of all $C^{k}$ cross-sections of the vector bundle ${\bf E}$ of $l(G)$-valued 2-forms on $M$.

	Let $F$ and ${\bf F}$ be manifolds on which $G$ acts on the left and such that $E = P\times_{G}F$ and ${\bf E} = P\times_{G}{\bf F}$, where $P$ is the total space of $P(M,G)$. In other words, $E$ is the quotient space of $P\times F$ by the group action. Similarly, ${\bf E}$ is the quotient space of $P\times {\bf F}$ by the action of the group $G$.

	To prove the following proposition we use the Atiyah-Singer index theorem. So, we are still assuming that $M$ and $G$ are compact.

	Therefore:
\begin{proposicao}
If a gauge field (a cross-section of ${\bf E}$) is associated to copied potentials that are locally gauge-equivalent, then there is: a non-trivial sub-group of $G$, denoted by $G'$; a $G'$-manifold $P'$; two $G'$-vector spaces $F'$ and $\bf F'$; and two elliptic partial
differential operators,
\begin{equation}
{\cal D}_{G}:C^{\infty}(P;P\times F)\rightarrow C^{\infty}(P;P\times
{\bf F})
\end{equation}
and
\begin{equation}
{\cal D}_{G'}:C^{\infty}(P';P'\times F')\rightarrow
C^{\infty}(P';P'\times {\bf F'})
\end{equation}
respectively $G$-invariant and $G'$-invariant, such that the
${\rm index}\; {\cal D}_{G'}$ can be defined as a nontrivial function of the ${\rm index}\; {\cal D}_{G}$.\label{main}
\end{proposicao}

	{\em Proof}: If a gauge field is associated to copied potentials that are locally gauge-equivalent, then such a field is reducible (Proposition \ref{Teo-Doria}). Therefore the bundle $P(M,G)$ is reducible (Definition \ref{def-redutibilidade}). So, there is a non-trivial sub-group $G'$ of $G$ and a monomorphism $\varphi:G'\rightarrow G$ such that one can define a reduced principal fiber bundle $P'(M,G')$ and a reduction $f:P'(M,G')\rightarrow P(M,G)$. Similarly we define $G'$-vector spaces $F'$ and $\bf F'$ and maps ${\bf f}:{\bf F'}\rightarrow {\bf F}$ and ${\rm f}: F'\rightarrow F$ such that:
\begin{equation}
{\rm f}(g'\xi') = \varphi(g'){\rm f}(\xi')
\end{equation}
and
\begin{equation}
{\bf f}(g'\zeta') = \varphi(g'){\bf f}(\zeta')
\end{equation}
for all $g'\in G'$, $\xi'\in F'$ and $\zeta'\in {\bf F'}$.

	Now consider $P\times F$ as the total space of the trivial vector bundle $P\times F\rightarrow P$, with a canonical projection. That vector bundle is noted $\chi$. Similarly the trivial vector bundles $P\times {\bf
F}\rightarrow P$, $P'\times F'\rightarrow P'$ and $P'\times {\bf F'}\rightarrow P'$, are respectively noted $\vartheta$, $\chi'$ and $\vartheta'$.

	Therefore, the diagram below commutes:
\[K_{G}(TP)\stackrel{\varphi^{*}}{\longrightarrow}K_{G'}(TP')\]
\[t_{{\rm ind}_{G}^{P}}\downarrow\;\;\;\;\;\;\downarrow t_{{\rm ind}_{G'}^{P'}}\]
\[R(G))\stackrel{\varphi^{*}}{\longrightarrow}R(G')\]
(Here $\varphi^{*}$ is induced by $\varphi$.)

	$\sigma({\cal D}_{G}) = \vartheta-\chi$ and $\sigma({\cal D}_{G'}) = \vartheta'-\chi'$. So, the homomorphisms $f$, $\bf f$ and f and the diagram given above induce the relation:
\begin{equation}
\sigma({\cal D}_{G'}) = \varphi^{*}(\sigma({\cal D}_{G}))
\end{equation}

	Thus, according to the diagram, $t_{{\rm ind}_{G'}^{P'}}(\sigma({\cal D}_{G'})) = \varphi^{*}(t_{{\rm ind}_{G}^{P}}(\sigma({\cal D}_{G})))$. If we use the Atiyah-Singer index theorem (\ref{Atiyah-Singer-D}), it can be noticed that\\
\begin{equation}
{\rm index}\;{\cal D}_{G'} = \varphi^{*}({\rm index}\;{\cal D}_{G}).\Box
\end{equation}

\begin{observacao}
{\rm The condition that ${\cal D}_{G}$ and ${\cal D}_{G'}$ are elliptic partial differential operators is necessary in our proof of proposition 3.1 in order to apply the Atiyah-Singer index theorem given by proposition 1.3.}
\end{observacao}

	The topological condition given in \cite{Doria-81}, in order to check whether there are false gauge field copies does not impose that the manifold $M$ should be compact, or that the Lie group must be compact (a very common situation in gauge theories). But when we apply the Atiyah-Singer index theorem to obtain the analytical condition for false copies, we must consider that both $M$ and $G$ are compact.$\Box$

\section{Conclusions}

	There are several points of contact between classical physics and $K$-theory: a method to prove the index theorem based on the asymptotics of the heat equation \cite{Atiyah-Calor} \cite{Gilkey}; and the physical interpretation of nonvanishing characteristic classes in terms of magnetic monopoles, solitons and instantons \cite{Mayer}. We present here a new $K$-theoretical result with consequence to physics: an analytical condition to the existence of `false' gauge field copies obtained from a topological condition.

	Our result has some limitations: it refers only to false copies; it is imposed that $M$ and $G$ are compact; and it is imposed that $G$ is semi-simple. We believe that it is possible to extend our results while eliminating those restrictions. (One possibility should be to modify the geometry of an irreducible principal fiber bundle in such a way as to handle true copies as false copies (see \cite{Costa-Amaral} on this). That is a possible way to define an analytical condition for generic copies through the use of the Atiyah-Singer index theorem. On the structure of $M$, we just recall that $K$-theory has a technique for dealing with locally compact spaces. Moreover, a possible way of generalizing our results to any Lie group (compact or non-compact) amounts (we hope) to the use of some kind of compactification \cite{Dikranjan}.

	Another side-result of the present work has to do with the categorical approach that underlies $K$-theory. When we formalize gauge theory and the gauge copy problem with tools from category theory we notice that our categorical formalization ends up as being too complicated when compared with the traditional approach based on set theory \cite{da Costa-92}.

	Moreover, we think it is possible to obtain the same result of Proposition \ref{main} without $K$-theoretical concepts. Firstly, these results make no reference to $K$-theoretical elements. The index theorem only appears in the proof of those propositions. Secondly, there is an analytical (non-topological) proof (non-topological) of the index theorem by the use of Zeta functions $\zeta(s) = \sum\lambda^{-s}$ ($\lambda$ denotes eigenvalues of an operator) \cite{Atiyah-Bott}. Therefore, we conjecture it may be possible to prove the analytical condition that we established for the existence of false gauge field copies in a similar way.

	Still another possibility has to do with proving our results with the help of the asymptotics of the heat equation \cite{Gilkey}, since that technique was also used to prove the index theorem. But those are questions to be answered in future papers.

	All those questions can be summed up in the following: the gauge field copy phenomenon is a remarkable feature of nonlinear gauge fields without a clear-cut physical interpretation. Copied fields belong to a nowhere dense, bifurcation-like domain in the space of all gauge fields in the usual topology \cite{Doria-84}. Copied fields are actually involved in a (possible) symmetry-breaking mechanism that remains strictly within the bounds of classical field theory \cite{Amaral}. Yet their contribution to path-integral computations remain unknown -- for instance, if Feynmann integrals are understood as Kurtzweil-Henstock (KH) integrals \cite{Muldowney} \cite{Peng-Yee}, the rather unusual topology of the space of all KH-integrable functions may lead to a new interpretation of the Feynman integrals over field and potential spaces in gauge theory, when given the adequate KH-structure.

	So, when we try to establish a connection between Atiyah-Singer theory and gauge copies, we are trying to expand the perspectives from where we can observe and try to understand the field copy phenomenon.

\section{Acknowledgments}

	The present work was partially written while A.S.S. was visiting Stanford University as a post-doctoral fellow. A.S.S. wishes to thank Professor Patrick Suppes for his hospitality. The authors acknowledge also support from CNPq, CAPES and Fapesp (Brazil). This paper was computer-formatted with the help of {\em Project Griffo} at Rio's Federal University.

%\newpage


\begin{thebibliography}{99}
\bibitem{Amaral} A.F. Furtado do Amaral, F.A. Doria and M. Gleiser, {\em Journal of Mathematical Physics}, {\bf 24}, 1888 (1983).
\bibitem{Atiyah-Bott} M.F. Atiyah, R.Bott, {\em Ann. of Math.} ${\bf 86}$, 374 (1967).
\bibitem{Atiyah-Calor} M.F. Atiyah, R.Bott, V.K.Patodi, {\em Invent. Math.} {\bf 19}, 279 (1973). See {\em Errata} in Invent. Math., {\bf 28}, 277 (1975).
\bibitem{Atiyah-Index-I} Atiyah,M.F., I.M. Singer, {\em Ann. of Math.} $\bf{87}$ (1968) 484-530.
\bibitem{da Costa-92} N. C. A. da Costa and F. A. Doria, ``Suppes Predicates for Classical Physics'', in J. Echeverr\'{\i}a et al., eds., {\em The Space of Mathematics}, Walter de Gruyter, Berlin-New York (1992).
\bibitem{Costa-Amaral} N. C. A. da Costa, F. A. Doria, A. F. Furtado do Amaral, and J. A. de Barros, {\em Found. Phys.} {\bf 24} 783 (1994).
\bibitem{Dikranjan} D. N. Dikranjan, I. R. Prodanov, L. N. Stoyanov, {\em Topological Groups - Characters, Dualities, and Minimal Group Topologies}, Marcel Dekker (1989).
\bibitem{Doria-81} F. A. Doria, {\em Comm. Math. Phys.} $\bf{79}$ 435 (1981).
\bibitem{Doria-84} F. A. Doria, in {\em Functional Analysis, Holomorphy and Approximation Theory II}, edited by G. I. Zapata. North-Holland, 69 (1984).
\bibitem{Doria-78} F. A. Doria and S. M. Abrah\~ao, {\em J. Math. Phys.} {\bf 19}, 1650 (1978).
\bibitem{Gilkey} P. B. Gilkey, {\em The Index Theorem and the Heat Equation}, Mathematical Lecture Series {\bf 4}, Publish or Perish (1974).
\bibitem{Kobayashi} S. Kobayashi and K. Nomizu, {\em Foundations of Differential Geometry}, I, John Wiley (1963).
\bibitem{Mayer} M. E. Mayer, {\em Gauge Theories, Vector
Bundles, and the Index Theorem}, Springer Verlag (1981).
\bibitem{Muldowney} P. Muldowney, {\em A General Theory of Integration in Function Spaces}, Longman (1987).
\bibitem{Peng-Yee} L. Peng-Yee, {\em Lanzhou Lectures on Henstock Integration}, World Scientific (1989).
\bibitem{Shanahan} P. Shanahan, {\em The Atiyah-Singer Index
Theorem, an Introduction}, Lecture Notes in Mathematics  {\bf 638} (1978).
\bibitem{Wu} T. T. Wu and C. N. Yang, Some Remarks About Unquantized Non-Abelian Gauge Fields, Phys. Rev. D {\bf 12}, 3843 (1975).
\end{thebibliography}
\end{document}